\documentclass{aa}  
\usepackage{graphicx}
\usepackage{txfonts}
\usepackage{longtable,ltcaption}
\usepackage{lscape}
\usepackage{natbib}
\usepackage{lscape}
\usepackage{amsmath}
\usepackage{wasysym}
\usepackage{graphics}
\usepackage{txfonts}
\usepackage{color}
\usepackage{times}
\usepackage{parskip}
\usepackage{pdflscape}
\usepackage{geometry}
\usepackage{marginnote}
\usepackage[table]{xcolor}
\usepackage{tabu}
\usepackage{stfloats}
\usepackage{mathtools} 
\usepackage{nicefrac}
\usepackage{float}

\makeatother

\newfont{\nlx}{cmssdc10 scaled 900}
\newfont{\mfont}{cmssdc10 scaled 760}
\definecolor{myblue1}{rgb}{0.0,0.604,0.831} 
\definecolor{myblue2}{rgb}{0.0,0.49,0.6745}
\definecolor{myblue3}{rgb}{0.0156,0.4078,0.9921}
\definecolor{myblue4}{rgb}{0.0,0.44,0.87}
\definecolor{myred1}{rgb}{0.529,0.019,0.017}
\definecolor{mycyan}{rgb}{0.63921569,0.0,0.48235294}

\newcommand{\brem}[1]{\textcolor{black}{\nlx #1}}

\newcommand{\Starlight}{{\sc Starlight}}

\def\zsun{$\mathrm{Z}_{\odot}$}
\def\rdisk{$\rm R_{\rm D}$}
\def\starlight{{\sc Starlight}}
\def\fado{{\sc FADO}}

\def\D4000{$D_{4000}$}

\def\rr{$\mathrm{R}^{\star}$}
\def\rbulge{$\mathrm{R}_{\rm B}$}
\def\rcons{$\mathrm{R}_{\rm C}$}
\newcommand{\sbb}{mag/$\sq\arcsec$}
\newcommand{\dmb}{$<\!\!\!\delta\mu_{9{\rm G}}\!\!\!>$}
\def\diskspec{$\mathrm{\hat{F}}_{\rm D}^{\star}(\lambda)$}
\def\diskspecN{$\mathrm{\hat{F}}_{\rm D}(\lambda)$}
\def\bulgespec{$\mathrm{F}_{\rm C}(\lambda)$}
\def\bulgespecFit{$\mathrm{F}_{\rm C}^{\star}(\lambda)$}

\def\sdssT{$T_{\rm SDSS}(\lambda)$}
\def\fd{$f_{\rm D}$}
\def\fs{$f_{\rm S}$}
\def\fc{$f_{\rm C}$}

\def\RY{${\cal RY}$}
\def\mstar{${\cal M}_{\star}$}

\def\mtmass{$\langle t_{\star,\textrm{B}} \rangle_{{\cal M}}$}
\def\mZmass{$\langle Z_{\star,\textrm{B}} \rangle_{{\cal M}}$}

\def\tsstar{$\Sigma_{\star}$}

\def\msolar{$\mathrm{M}_{\odot}$}
\def\zsolar{$\mathrm{Z}_{\odot}$}

\newfont{\hvss}{cmssdc10 scaled 1540}
\def\?{{\bf\color{red}?}}
\def\reff{$\mathrm{R}_{\rm eff}$}

\begin{document} 

\title{Indications of the invalidity of the exponentiality of the disk within bulges of spiral galaxies}
   \author{
          Iris Breda
         \inst{\ref{IA-CAUP},\ref{UPorto},\ref{IA-FCiencias}}  
          \and
          Polychronis Papaderos
          \inst{\ref{IA-FCiencias},\ref{IA-CAUP}}
          \and
          Jean Michel Gomes
          \inst{\ref{IA-CAUP}}
          }
\institute{Instituto de Astrof\'{i}sica e Ci\^{e}ncias do Espaço - Centro de Astrof\'isica da Universidade do Porto, Rua das Estrelas, 4150-762 Porto, Portugal \label{IA-CAUP}
         \and
Departamento de F\'isica e Astronomia, Faculdade de Ci\^encias, Universidade do Porto, Rua do Campo Alegre, 4169-007 Porto, Portugal \label{UPorto}
         \and
Instituto de Astrofísica e Ciências do Espaço, Universidade de Lisboa, OAL, Tapada da Ajuda, PT1349-018 Lisboa, Portugal 
\label{IA-FCiencias}
\\
             \email{iris.breda@astro.up.pt}
             }

   \date{Received ???; accepted ???}

\abstract
{A fundamental subject in Extragalactic Astronomy concerns the formation and evolution of late-type galaxies (LTGs). The standard scenario envisages a two-phase build-up for these systems, comprising the early assembly of the bulge followed by disk accretion. However, recent observational evidence points to a joint formation and perpetual co-evolution of these structural components. Our current knowledge on the properties of bulge and disk is, to a large degree, founded on photometric decomposition studies, which sensitively depend on the adopted methodology and enclosed assumptions on the structure of LTGs. A critical assumption whose validity was never questioned is that galactic disks conserve their exponential nature up to the galactic center. This, although seemingly plausible, implies that bulge and disk co-exist without significant dynamical interaction and mass exchange over nearly the whole Hubble time.
}
{Our goal is to examine the validity of the standard assumption that galactic disks preserve their exponential intensity profile inside the bulge radius (\rbulge) all the way to the galactic center, as generally assumed in photometric decomposition studies.
}
{We developed a spectrophotometric bulge-disk decomposition technique that provides an estimation for the net (i.e. disk-subtracted) spectrum
of the bulge. Starting from an integral field spectroscopy (IFS) data cube, this tool computes the integrated spectrum of the bulge and disk, scales the latter considering the light fraction estimated from photometric decomposition techniques, and subtract it from the former, thereby allowing for the extraction of the net-bulge spectrum. Considering that the latter depends on the underlying assumption for the disk luminosity profile, a check of its physical plausibility (e.g., positiveness and spectral slope) places indirect constraints on the validity of the disk's assumed profile inside the radius $\rm R^{\star} <$ \rbulge. In this pilot study we tested three different disk configurations (the standard exponential disk profile, and a centrally flattened or down-bending exponential disk profile).
}
{A systematic application of our spectrophotometric bulge-disk decomposition tool to a representative sample of 135 local LTGs from the CALIFA Survey yields a significant fraction (up to $\sim$ 30 (20)\%) of unphysical net-bulge spectra when a purely exponential (centrally flattened) intensity profile is assumed for the disk. This never occurs for disk's profiles involving a centrally decreasing intensity.
}
{The obtained results suggest that, for a significant fraction of LTGs, the disk component shows a down-bending beneath the bulge. If proven to be true, such result will call for a substantial revision of structural decomposition studies for LTGs and have far-reaching implications in our understanding of the photometric properties of their bulges. Given its major relevance, it appears worthwhile to further explore the central stellar surface density of galactic disks through an improved version of the spectrophotometric decomposition tool here presented, and its application combining deep surface photometry, spatially resolved spectral synthesis and kinematical analyses.
}

\keywords{galaxies: spiral -- galaxies: bulges -- galaxies: evolution}
\maketitle

\parskip = \baselineskip

\section{Introduction \label{intro}}


One of the greatest challenges in Extragalactic Astronomy is to fundamentally explain the formation and evolution of late-type galaxies (LTGs) and their main structural components. To this end, it is crucial to assess the build-up histories of galactic disks and bulges, given that the formation of the latter is intimately linked to the genesis and growth of supermassive black holes (SMBHs) and their regulatory role on galaxy evolution \citep[e.g.,][]{KorHo13,HecBes14,MarNav18}.

%

Traditionally, galactic disks were thought to form early-on via violent quasi-monolithic gas collapse \citep{Lar74} or galaxy mergers \citep[e.g.,][]{BarHer96,SprHer05,BouJogCom05} and/or to gradually assemble around pre-existing monolithically collapsed bulges \citep[e.g.,][]{Kau93,Zoc06}.
Lately, however, thanks to the recent development of observing facilities and improved modeling techniques that employ increasingly sophisticated physical recipes, it became apparent that LTGs undergo rather complex assembly histories, comparatively to what was previously envisaged.

On the other hand, the literature regarding formation and evolution of LTGs reports contradictory results, leading to conflicting interpretations. For instance, by performing disk-to-bulge decomposition for 180 galaxies from the 3DHST Legacy Survey with redshift 1.5 < $z$ < 4.0, \cite{Sac19} find that from the redshift interval z > 2 to z < 2, the scale-length of two-component galaxies undergoes an increase ($\sim$1.3 times) whereas their bulge sizes and bulge/total ratio (B/T) remain almost constant. The authors infer that z$\sim$2 is mostly a disk formation period while bulges had formed at earlier times (z > 2).\footnote{This interpretation, however, seems to be in conflict with the fact that significant disk growth together with a constant B/T must also imply bulge growth.}
This conclusion has been recurrently reported in the literature \citep[e.g.,][]{Mar16,Mar17}, usually resulting from photometric bulge-disk decomposition studies. In contrast, other works where different techniques were applied suggest instead that bulges and disks grow and evolve jointly, opposing the prevailing view of two independent formation scenarios for bulge and disk build-up. In a study by \cite{vD13}, by applying the abundance matching technique to Milky Way (MW) progenitor candidates out to z = 2.5, it is found that MW-like LTGs have built $\sim$90\% of their present stellar mass (\mstar) after z = 2.5 with the star-formation peak  occurring before z = 1. Additionally, they show that the bulge buildup was prolonged, occurring between 1 < $z$ < 2.5. In this period, the mass in the central 2 kpc of MW progenitors increases by a factor of $\sim$3, ruling out models in which bulges were fully assembled first and disks gradually formed around them.
In addition, a recent study by \citet[][hereafter BP18]{BrePap18} where a representative sample of 135 local LTGs\footnote{The present study is the result of an extended analysis applied to the same galaxy sample as in BP18.} was analyzed by combining three techniques, namely surface photometry, spatially resolved spectral modeling and post-processing with RemoveYoung (\RY) \citep{GP16-RY}, demonstrates that LTG bulges form a $continuous$ $sequence$ with regard to their \dmb\ \footnote{\dmb\ (mag) is defined by BP18 as the difference between the mean $r$ band surface brightness of the  present-day stellar component and that of stars older than 9 Gyr.}, mean stellar age and metallicity (\mtmass\ and \mZmass, respectively), across $\sim$3 dex in log(\mstar) and $>$ 1 dex in log of stellar surface density (\tsstar).
Moreover, they find that physical properties of bulges and their parent disks are linked, pointing once more to a joint evolution between bulge and disk. 
Even though they adopt significantly different methodologies and galaxy samples, the two aforementioned studies find strong evidence for a unified formation scenario of LTGs, where bulge build-up time-scale and mass (total and relative) are dictated by the total galaxy mass \citep[see also][]{Gan07}. It is worth stressing that these studies are not based on structural decomposition techniques, being therefore free from prior assumptions on the LTG's individual stellar components (bulge, disk, bar) that might impact the obtained results.

Concerning the disk of LTGs, these complex stellar structures are mainly characterized by a radial light distribution $\rm I(R)$ that is usually well fit by an exponential law with the generic form $\rm I(R) \propto e^{-R/\alpha}$, where $\alpha$ is the disk scale-length. As a result, galaxy disks were initially thought to universally follow an exponential decay in their surface brightness profiles (type~I) \citep{Vau59}, yet subsequent work has revealed the existence of disk galaxies that deviate from the exponential law in their outskirts, exhibiting down/up-bending surface brightness profiles (SBPs) \citep[type~II and type~III, respectively,][]{Fre70}. These observations prompted a revision of the proposed theories for the formation of such stellar structures and accurate modeling of possible divergences from the exponentiality \citep[e.g., ][]{Poh08,Lai14,Wat19}.
The last decades were crucial for the development of this field of research, due to the overall improvement of computational power and tools. 
\citet{Dal97} tries to explain the range of observed disk properties by using a set of gravitationally self-consistent models for disk collapse, assuming that while collapsing, the resulting dark matter halo has a universal density profile, the angular momentum is conserved and its initial distribution is comparable to that produced by an external tidal torque.
Their work predicts that several disk properties are intimately related, such as total mass and initial angular momentum -- the collapse of a gas cloud with low initial angular momentum will give rise to a high mass, high surface brightness galaxy whereas the opposite is valid in the case of a gas cloud with high initial angular momentum. 
The shape of the rotation curve also appears to be tightly connected with the initial angular momentum. Low angular momentum disks (generally higher mass galaxies) are centrally concentrated being globally unstable to non-axisymmetric perturbations which may result in angular momentum transfer and secular bulge/bar formation. This contrasts with high angular momentum disks (usually lower mass galaxies) which display a slowly rising rotation  curve that is not so prone to instabilities (i.e., precursors of bulge/bar formation).
As for observational studies, a recent IFS study by \cite{RL17a} attributes deviations from an exponential slope to radial stellar migration, proposing it to be more efficient in type~III disks as compared to type~II. A supplementary study by these authors adds further support to this scenario through the modeling of Milky Way-mass disks in cosmological simulations \citep{RL17b}. 

Whereas significant progress has been achieved in our understanding of disks in their outer parts, the same does not apply for their central regions. A critical question that has been barely explored is whether the exponential slope of the disk is valid all the way to the LTG center, i.e., beneath the bulge. Although SBPs of bulgeless\footnote{In this work, \emph{bulgeless galaxies} denote pure disk galaxies without evidence for a central luminosity excess.} galaxies show a pure exponential luminosity profile \citep[e.g.,][]{SacSah16}, the central radial distribution of the disk for the majority of LTGs remains unexplored. Considering that the existence of a bulge prevents direct observation of the disk in the inner region of the galaxy, it is standard procedure to preserve the assumption that the disk follows an exponential profile.


The implications of this simplifying assumption are manyfold and fundamental: this fitting approach actually presumes that bulge and disk co-exist without significant dynamical interaction and mass exchange (e.g., stellar migration, kinematical heating) over several Gyr of galactic evolution, which appears to be in contrast with conclusions from previous studies that advocate a joint evolution of these two galaxy entities.
Furthermore, in the light of the assumption that disks gradually assembled around bulges that were formed prior to and independently from the disk, the most reasonable expectation should be a decrease in the disk's stellar surface density in the innermost part of the galaxy, i.e., the opposite of what it is commonly assumed. If bulges are a product of monolithic collapse, it appears legitimate to consider that the inherent high gas and stellar velocity dispersion ($\sigma_{\star}$) in the bulge would act against the build-up of a dynamically cold disk inside the bulge radius.

Another implicit assumption enclosed within the generally adopted exponentiality of the disk to the LTG center is the absence of significant 
stellar age or metallicity gradients within the bulge radius, as well as that the specific star-formation rate (sSFR) is nearly invariant throughout the disk. Recently, IFS observations allowed a spatially resolved exploration of stellar populations of galaxies across a significant part of their radial extent (up to 1.5 to 2~ effective radii, $\rm R_{eff}$). Many studies demonstrate that some LTGs present age and metallicity gradients across their disks \citep[see, e.g.,][]{Gon15,God17} pointing to the non-negligible effect of a non-uniform sSFR throughout the galaxy. In addition, a systematic analysis of the radial profiles of mass and luminosity weighted ages of the same LTG sample used in this study \citep{Bre20a} reveals quite significant stellar age gradients within the bulge radius, whose slope (positive/negative) is anti-correlated with total galaxy mass.

From the theoretical viewpoint, in the past decades there has been some evidence supporting the non-preservation of the exponential profile of the disk within the bulge. 
Attempts to model the rotation velocity profile or the stellar surface density of the disk component by means of semi-analytic models \citep{KuiDub95}, N-body simulations \citep{WidDub05} or, more recently, hydrodynamical simulations \citep{Obr13} show that, in the presence of a bulge (whose stars are characterized by significantly higher $\sigma_{\star}$ as compared to the disk), the orbits of disk stars (mostly supported by rotational velocity, V$_{\rm rot}$) are gradually kinematically heated by cumulative weak interactions with the bulge. 
According to these results, interaction between both collisionless stellar populations over several Gyr will eventually lead to a scarcity (or even depletion) of rotation-dominated stellar populations in the galactic center, resulting in a flattening or even a sharp central decrease of the disk's particle density, as shown in Fig.~2 of 
\citet{Obr13}. In this context, it is noteworthy that from the observational point of view, the need for a central flattening of exponential profiles in many dwarf galaxies was established through surface photometry of early-type and late-type systems \citep{BinCam93,P96a,Noeske03}. 

If the assumption on the inner disk's exponential nature is proved to be incorrect, important implications might be expected for structural studies of LTGs. For instance, while performing 1D surface photometry the standard procedure for the photometric decomposition of such galaxies involves the determination and subtraction of the disk contribution by approximating its SBP by an exponential function. The residual central luminosity excess is attributed to the bulge and fitted with a S\'ersic model, the best-fitting parameters of which ($\eta$, $\rm R_{eff}$, radial extent, total magnitude) are used for bulge classification into classical bulges (CB) and pseudo-bulges (PB). 
Even in the case where all the structural components such as bulge and disk are fitted simultaneously, as commonly occurs with 2D surface photometry codes such as IMFIT \citep{Erwin2015} or GALFIT \citep{Peng2010}, by assuming an incorrect surface brightness distribution for the disk one would still introduce a systematic bias.
A possible overestimation of the disk luminosity inside the bulge radius due to the false assumption that it retains its exponential slope all the way to the center, would impact determinations of the luminosity and structural properties of the bulge. This might offer an explanation for the relatively weak correlation between $\eta$ and bulge magnitude \citep[see for e.g.,][]{Hea14,Mos14}.

As a pilot attempt to investigate the disk's radial distribution within the bulge and evaluate the validity of the conventional and universal assumption of extrapolating its exponential intensity profile, here we present a study where we test this hypothesis in the context of spectral analysis and modeling. To this end, we developed a tool that allows us to estimate the net spectral energy distribution (SED) of the bulge after removal of the disk contribution using a combined photometric and spectral modeling approach. The tool was applied to a representative sample of the local LTGs population, comprising 135 galaxies from the CALIFA IFS survey \citep{Sanchez12-DR1,Sanchez16-DR3}.
It is worth mentioning that there were previous attempts to perform spectrophotometric decomposition of galaxies based on IFS data \citep{Men19,John17} and long-slit spectroscopy \citep{Johnston12,Johnston14,Sil12}. However, the methodologies adopted in these studies greatly differ from the one presented here, with the most significant difference being that these authors fix the surface brightness distribution of the disk to the standard exponential law. 
Consequently, these studies do not explore, per design, possible deviations from the exponentiality of the disk in its central part; the exponentiality of the disk was explicitly assumed. Additionally, these tools were applied to early-type (ETGs) and lenticular (S0s) galaxies, respectively, i.e. galaxies with nearly homogeneous stellar populations in terms of age, contrary to LTGs.

Section sample describes the sample selection, Sect. \ref{meth} outlines the adopted methodology, Sect. \ref{resDiskSub} presents the main results and finally Sect. \ref{conc} is dedicated to the discussion of the obtained results and summarizing the conclusions.

\section{Sample description \label{sample}}

The galaxy sample here analyzed was selected from the 3$^{\rm rd}$ Data Release of the CALIFA integral field spectroscopy (IFS) survey \citep[667 galaxies;][see http://califa.caha.es]{Sanchez16-DR3}. It is constituted by 135 non-interacting, nearly face-on local ($\leq$130 Mpc) LTGs (see BP18 for a complete description of sample) and it was assembled aiming for high representativity of the LTG population in the local Universe, spanning a range of $\sim$3 dex in log(\mstar) and $>$ 1 dex in log(\tsstar). In BP18 the galaxy sample was tentatively subdivided into three $< \mkern-6mu \delta\mu_{9{\rm G}} \mkern-6mu >$ intervals. 
This quantity was there defined as the difference $\mu_{\rm 0\,Gyr}$-$\mu_{\rm 9\,Gyr}$ between the mean $r$ band surface brightness of the  present-day stellar component and that of stars older than 9 Gyr (a \dmb\ $\approx$ 0 mag implies that the bulge has completed its buildup earlier than 9 Gyr ago ($z\simeq 1.34$) while a \dmb\ of --2.5 mag denotes a contribution of 90\% from stars younger than 9 Gyr). Interval A (\brem{iA}; \dmb\ $\leq$ --2.5 mag; 34 galaxies) includes the least massive galaxies which host low-mass, young bulges with low stellar metallicity, frequently classified as star-forming (SF) after the BPT spectroscopic classification scheme \citep{BalPhiTer81}. In contrast, interval C (\brem{iC}; \dmb\ $\geq$ --0.5 mag; 43 galaxies) contains the most massive galaxies with the most massive, dense, old and chemically enriched bulges, typically falling in the class of AGN/LINER. LTGs that fall within interval B (\brem{iB};  --1.5 mag to --0.5 mag; 58 galaxies) display intermediate characteristics in all measured properties. As for the frequency of bars in our sample they represent about 40\% ($\sim$1/3 for \brem{iA}, $\sim$1/3 for \brem{iB} and $\sim$2/3 for \brem{iC}). This subdivision and subsequent examination of bulge and galaxy properties within each interval led us to conclude that the galaxy total mass is the main evolutionary driver for LTGs, being tightly connected with the bulge's stellar mass and surface density, mean stellar age and metallicity, current photo-ionization mechanism and mean stellar age and metallicity of the disk.

\section{Data analysis \label{meth}}

The concept used here consists of assuming different intensity profiles for the innermost part of the disk, subtract their contributions from the total central luminosity and subsequently fit the remaining (net) bulge spectra, in order to assess whether they are physically plausible. An unphysical or implausible SED for the disk-subtracted bulge implies that the underlying assumption of the exponentiality of the disk is invalid.

\begin{figure*}[t]
\centering
\includegraphics[width=1\linewidth]{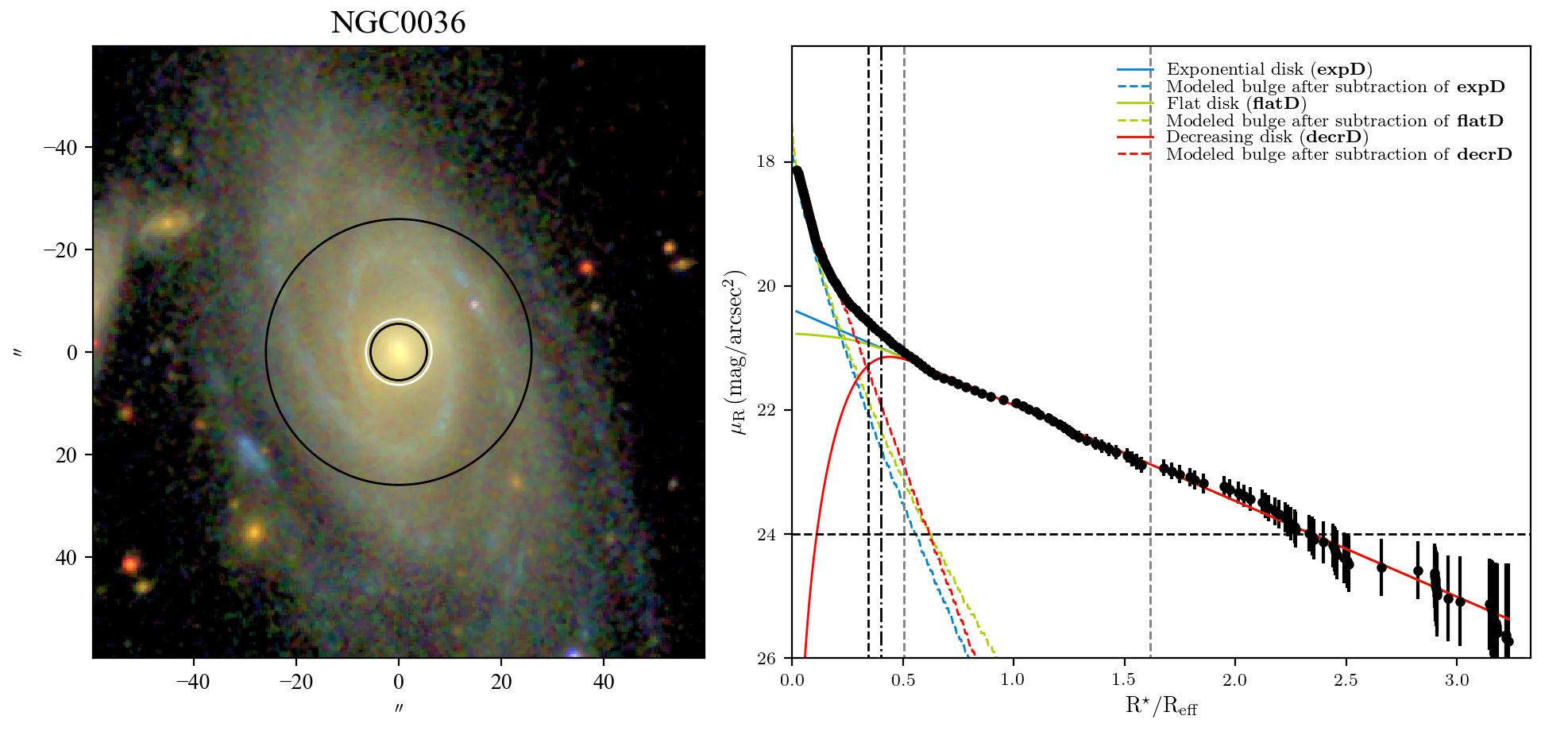}
\caption{Left: SDSS true-color image of LTG \object{NGC 0036}. The white circle overlaid in the SDSS image depicts the bulge diameter (with radius \rbulge) and the innermost black circle illustrates the bulge radius that was considered for this study (with radius \rcons, defined in Sect.~\ref{spec}). The radius of the outermost black circle corresponds to the limit of the disk (with radius \rdisk)\protect\footnotemark. Right: SDSS $g$-band SBP illustrating the assumed three disk configurations (black points -- observed profile; blue solid line -- standard exponential disk, \brem{expD}; green solid line -- inwardly flattening disk, \brem{flatD}; red solid line -- centrally depressed disk, \brem{decrD}). The colored dashed lines represent the modeled bulge after subtraction of the respective disk's profile. The horizontal line denotes the limiting surface brightness at $\mu$ = 24~\sbb. Grey vertical lines denote the considered limits for the disk (the outermost grey line is illustrated by the uttermost circle in the l.h.s), dashed vertical black line depicts \rbulge\ (innermost black circle at the l.h.s) and the dashed-dot vertical black line the \rcons\ (white circle at the l.h.s). The galactocentric distance (x axis) is normalized to \reff. }
\label{disks}
\end{figure*}

The adopted methodology combines modeling of binned IFS spectra -- by means of two population spectral synthesis (PSS) codes: \Starlight\ \citep{Cid05} \& \fado\ \citep{GomPap17} -- with surface photometry of optical images. Surface photometry was carried out on SDSS $r$- and $g$-band images with the goal of estimating the expected light fraction of the disk inside the bulge radius when assuming different radial intensity distributions. As shown in Fig.~\ref{disks}, we assumed three profiles for the radial intensity of the disk within the bulge radius, \rbulge: a) purely exponential, as commonly assumed b) an inwardly flattening and c) a centrally depressed, so that at the galactic center its contribution is virtually zero. Subsequent integration of the different light growth curves allowed to determine the fraction of light within \rbulge\ pertaining to the disk (\fd), for each disk configuration. 

In parallel, we conducted a spectral fitting analysis: after having a clear-cut definition of \rbulge\ (see Sect.~\ref{3disk}) and the disk radial extent, individual spaxels were integrated into one spectrum for the respective stellar components. For each galaxy, bulge and disk spectra were modeled by both \starlight\ \& \fado\ using two simple stellar population (SSP) spectral libraries -- Z4, comprising SSPs from \citet{BruCha03} for 38 ages between 1~Myr and 13~Gyr for four stellar metallicities (0.05, 0.2, 0.4 and 1.0~$Z_{\odot}$), referring to a Salpeter IMF \citep{SalIMF} and Padova 2000 tracks \citep{Gir00}, and Z5 which is identical to Z4 in terms of age coverage except for being supplemented by SSPs with a metallicity of 1.5~\zsun. The use of two stellar libraries and two conceptually distinct spectral fitting codes permitted to uncover to what extent the obtained results depend on the spectral modeling technique.

The best-fitting synthetic stellar spectrum of the normalized SED of the disk was then scaled according to the previously determined \fd\ (after correction from the SDSS $r$- or $g$-band transmission curve) and subsequently subtracted from the bulge best-fitting synthetic stellar SED, this way theoretically obtaining the net-bulge SED. Finally, we assessed the soundness of the obtained spectra by refitting them with both PSS codes. 
Considering that this approach consists in the application of spatially resolved spectral modeling on IFS data, it is, therefore, free from any prior assumption on its stellar populations and star formation histories (SFH) throughout the galaxy.

\footnotetext{The circle depicting \rdisk\ does not precisely correspond to the analyzed area, being instead the equivalent radius (the radius of a circle with an area equal to the sum of all spaxels that belong to the disk). The l.h.s of Fig.~\ref{maps} displays the spaxels that were considered in constructing the average spectrum of the disk for \object{NGC 0036}.}


\subsection{Photometric decomposition assuming three different radial profiles for the disk}\label{3disk}

To test the standard assumption that the disk conserves its exponential nature all the way to the galactic center, we resort to structural analysis which permits to decompose galaxies into their main stellar constituents. Seeking for a uniform, clear-cut definition for the bulge radius \rbulge, without relying on strong prior assumptions on the photometric structure of LTGs, this quantity was determined by fitting a single S\'ersic model to the central luminosity peak. It was determined at an extinction-corrected surface brightness level $\mu_{\rm lim}$ of 24 \sbb, this way encircling nearly all the flux from the bulge. For this purpose we used our 1D surface photometry code iFIT \citep{ifit}\footnote{As discussed in BP18 in further detail, we additionally performed full image decomposition with iFIT, IMFIT and GALFIT. We generally found a minor dependence of \rbulge\ on different codes and profile fitting schemes.}, after visual inspection of the morphology and $g$--$i$ color maps.
As for the disk, it was modeled by fitting the standard exponential model ($\mathrm{I}_{e}$) at intermediate radii (see BP18 for additional details on the photometric analysis), which equation in intensity units is given by:

\begin{equation}
\mathrm{I}_{e}(\rm R^{\star}) = I_{0} \cdot e^{\rm - R^{\star}/\alpha},
\end{equation}

where I$_{0}$ is the intensity at the galactic center, $\rm R^{\star}$ the radius (i.e., distance from the center) and $\alpha$ is the scale-length, in $\arcsec$.

After estimating the best-fitting parameters for the exponential disk for each galaxy in the sample, we adapted the previously estimated disk to a centrally flattened and a down-bending surface brightness profile within \rbulge by means of the modified exponential ($\mathrm{I}_{\hat{e}}$) distribution proposed by \citet{P96a}:

\begin{equation}
\renewcommand*{\arraystretch}{1.5}
\begin{array}{l}
\mathrm{I}_{\hat{e}}(\mathrm{R}^{\star}) = \mathrm{I}_{e}[1 - \epsilon_{1} e^{\mathrm{P}_{3}(\rm R^{\star})}] \\
\rm P_{3}(R^{\star}) = \left( \dfrac{\rm R^{\star}}{R_{core}} \right)^{3} + \left( \dfrac{\rm R^{\star}(1-\epsilon_{1})}{\alpha \epsilon_{1}} \right)
\end{array}
\renewcommand*{\arraystretch}{1}
\end{equation}


In all cases, we set the core radius $\rm R_{core}$ equal to \rbulge\ so that $\mathrm{I}/\mathrm{I}_{e}$ starts diverging from 1 at this radius (i.e., decreasing the disks's intensity relative to that corresponding to an exponential fitting law).
As for the central intensity depression, $\epsilon_{1} = \Delta \rm I / I_{0}$, we tested two different cases: an intermediate case characterized by a nearly constant intensity within \rbulge\ (adopted $\epsilon_{1} =\alpha/1.5 \rm R_{core}$) and a steeply down-bending disk profile reaching zero intensity ($\epsilon_{1} = 1$) at the center ($\rm R^{\star}=0$) of the disk. These two modified exponential profiles are referred as \brem{flatD} and \brem{decrD}, respectively, and are illustrated in the r.h.s. panel of Fig.~\ref{disks}, in addition to the standard exponential disk \brem{expD}.

Although a full photometric analysis and characterization of the different bulge luminosity profiles that remain after subtraction of the assumed disks is out of the scope of this article, as a sanity check we used a simple $\chi^2$ minimization algorithm to fit a S\'ersic model to the residuals, as a tentative assessment of the plausibility of the obtained bulge profiles.
We document values for the S\'ersic index within $0.3 \leq \eta \leq 2.3$ (not reaching, in any circumstance, unfeasible values such as 0.2 > $\eta$ > 8) and an average absolute difference of 0.5 between the best-fitting $\eta$ for \brem{expD} and \brem{flatD} and 0.2 for \brem{expD} and \brem{decrD} (see Fig.\ref{disks} for a visual representation of the three possible bulge luminosity profiles for \object{NGC 0036}). Regarding the obtained estimates for the B/T, we report values within $3\% \leq \mathrm{B/T} \leq 51\%$ and an average increase of $\sim 15\%$ between the B/T estimated after subtraction of \brem{expD} and the same after subtraction of \brem{flatD} and $\sim 25\%$ when comparing \brem{expD} with \brem{decrD}. Visual inspection of the remaining excess and respective models suggests that all the obtained bulge luminosity profiles are physically reasonable, demonstrating that experiments involving surface photometry do not yield strong discriminators of deviations of the disk from the exponentiality.


\subsection{Construction of bulge and disk spectra \label{spec}}

\begin{figure*}[h!]
\centering
\includegraphics[width=1\linewidth]{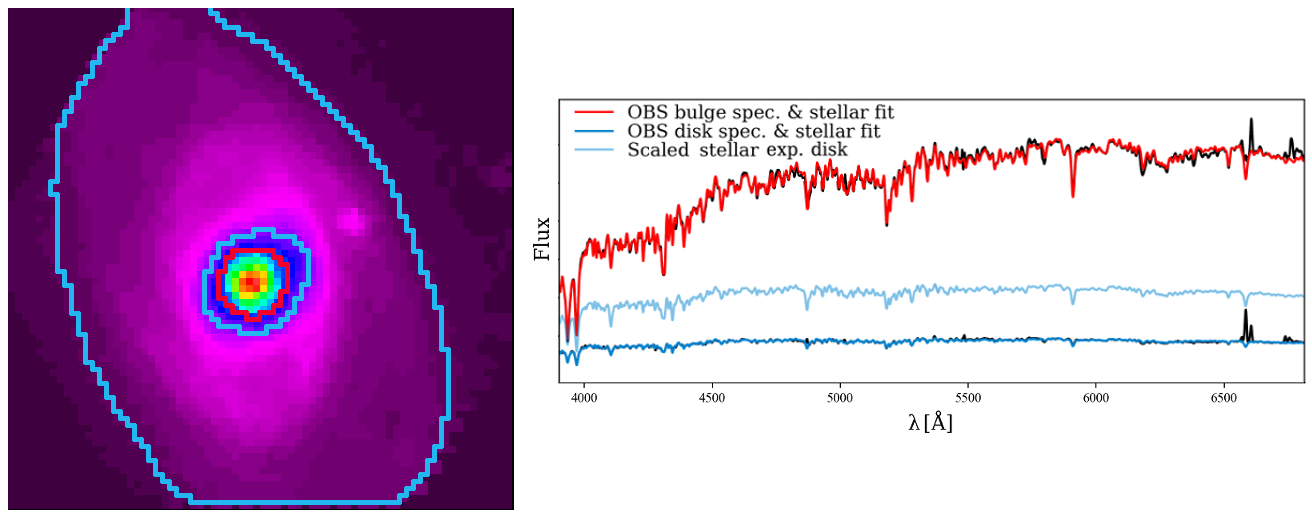}
\caption{Illustration of the CALIFA IFS data (spaxel-by-spaxel map) of  \object{NGC 0036} (l.h.s.) where it is shown the emission-line-free pseudo-continuum between 6390 and 6490~\AA. Within the red contour are the spaxels that were integrated to obtain a single bulge spectrum which is shown in the r.h.s along with its respective best-fitting stellar SED as obtained by \starlight, Z4, in red. As for the average disk spectrum displayed in the r.h.s overplotted by its best-fitting stellar spectrum, in dark-blue, was obtained by integrating the spaxels that lay between the two blue contours in the l.h.s. and subsequent division by the considered number of spaxels. Additionally, it is plotted in light-blue at the r.h.s the best-fitting stellar SED for the disk after scaling according to the exponential model.}
\label{maps}
\end{figure*}

By integrating IFS CALIFA data (after correction of spectra
in individual spaxels for intrinsic stellar motions) we constructed two spectra per galaxy: a summed up spectrum of the bulge and an average spectrum of the disk, normalized to one spaxel (1 $\sq\arcsec$ given that CALIFA data-cubes have a pixel scale of 1 spaxel). Bearing in mind that this is a pilot study and considering that the spectroscopic modeling and subsequent subtraction of the bar contribution is a non-trivial task, there was no attempt to model separately the bar component. Additionally, by computing an average disk spectrum from a significant number of spaxels (see Fig. \ref{maps}), whereas the bar is confined to a smaller number of spaxels as compared to the disk in any barred galaxy, it is expected that the spectroscopic contribution from the bar is smoothed out in the final disk spectrum, so that the bar will only marginally contaminate the average spectrum for the disk (i.e., the region outside \rbulge).

Using an adaptation of the isophotal annuli (\brem{isan}) surface photometry technique by \citet{P02}, which consists on the computation of of the mean surface brightness for a given filter within logarithmically equidistant isophotal zones obtained from a reference image, in this case the emission-line-free pseudo-continuum between 6390 $\AA$ and 6490 $\AA$, all galaxies of the sample were previously segmented into 18 isophotal zones.
Having the sample galaxies already segmented and the information on the radial extent of the bulge, the following procedure was to determine the zones that lie within the bulge and the ones within the disk, as pictured in Fig.~\ref{maps}. Here we defined a more conservative radial extent of the bulge (\rcons, see black circle overlaid with SDSS true color image at the left panel and dashed vertical line at the right panel of Fig.~\ref{disks}) which is the galactocentric radius of the last zone that lies within \rbulge, as given by the structural analysis (mean \rcons\ $\sim$ 0.87 $\cdot$ \rbulge; $\sigma$ = 0.13 for the LTG sample -- given the minor difference one can assume that \rbulge\ $\simeq$ \rcons).
\\

The l.h.s of Fig.~\ref{maps} illustrates the spaxels that were used to construct the two spectra in the case of \object{NGC 0036}. The total spectrum within the bulge region \bulgespec, (i.e., bulge plus a possible disk contribution, displayed in black at the r.h.s of Fig.~\ref{maps}, with its best-fitting stellar SED \bulgespecFit, overplotted in red) was created by summing up the spaxels residing within the red contour, i.e., all the spaxels that pertain to the zones within \rcons. The normalized disk spectrum \diskspecN, (plotted in the r.h.s of the same Fig. in black, with its best-fitting stellar SED \diskspec, overplotted in dark-blue) was constructed by summing up the spaxels that occupy the locus between the two blue contours (i.e., between one zone after the last one pertaining in the bulge and zone 14\footnote{We decided to exclude the last 4 zones from the analysis due to the decreasing surface brightness $\mu$, and consequently decreasing signal-to-noise ratio, of the outermost spaxels. Throughout the sample, the 14$^{th}$ zone has an average value of $\mu$ = 23.6 \sbb.}), subsequently dividing by the considered number of spaxels. Still at the r.h.s of Fig.~\ref{maps}, it is plotted in light-blue the best-fitting stellar SED for the disk, after scaling according to \brem{expD}.

Considering that most of our disks host significant star-formation which manifests itself through strong emission lines, by directly subtracting the observed disk spectra one would get artificially deep absorption features in the net-bulge spectrum. A way of avoiding this is to conduct first the spectral fitting of bulge and disk, which will result in emission-line free spectra \diskspec\ \& \bulgespecFit. To this end, spectral modeling of the disk and bulge spectra was carried out using the PSS codes \starlight\ \& \fado\ in the spectral range between 3900 and 6800~\AA\ adopting the Z4 and Z5 libraries (cf. Sect. 2.2). The four spectral modeling runs will be later referred to as SLZ4, SLZ5, FDZ4, FDZ5, respectively.
For \starlight, a purely stellar code, strong emission lines were masked out before fitting while for \fado\ these are used to achieve self-consistency between the nebular and the stellar emission, while deriving the SFHs.

\begin{figure*}[b] 
\centering
\includegraphics[width=0.7\linewidth]{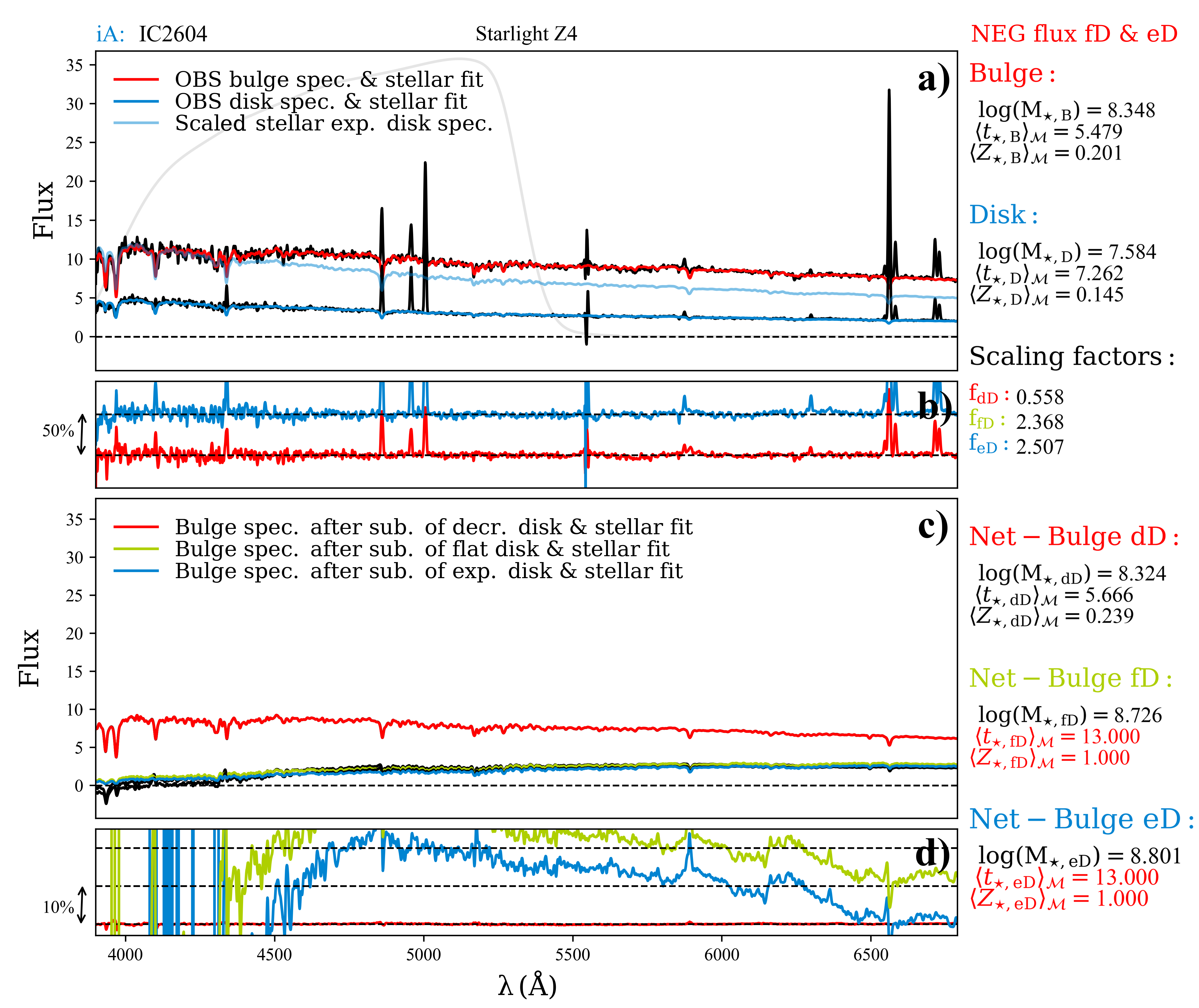}
\caption{The resulting spectra and respective fits for LTG \object{IC 2604} -- panel \textbf{a)} displays the spectrum of the bulge (red), disk (dark-blue) and scaled disk (light-blue) according to the exponential disk distribution (additionally it can be seen the overplotted SDSS $g$-band transmission curve in soft grey). Panel \textbf{b)} displays the residuals (in $\%$) between the modeled and observed spectrum for the bulge (red) and disk (dark-blue) with the black horizontal line corresponding in either case to a percentage deviation of 0$\%$. The vertical arrow corresponds to a percentage deviation of 50$\%$. At panel \textbf{c)} it is shown in black the resulting net-bulge spectra after subtraction of the three different models for the disk and overplotted are the respective stellar fits. The residuals, i.e., difference between obtained net-bulge SED and respective stellar fits divided by the observed, are shown in panel \textbf{d)}. As in panel \textbf{b)} we shift the residuals  by an arbitrary amount (in this case, by 10$\%$; cf. vertical arrow) for the sake of better visibility. Labels on the r.h.s of the panels list the bulge and disk stellar mass $ \log({\cal M}_{\star})$, and mass-weighted mean stellar age \mtmass\ and metallicity \mZmass\ prior to subtraction, the scaling factors for each of the disk configuration and \mtmass\ \& \mZmass\ for each of the fits (where dD, fD and eD corresponds to centrally decreasing, flat and exponential disk, respectively), as obtained with \starlight\ and Z4 stellar library.}
\label{disk_sub_im} 
\end{figure*}

\subsection{Scaling of \diskspec:}

Subsequently, we estimated how much light within \rbulge\ belongs to the disk component and, accordingly, how much it is needed to scale \diskspec\ to ensure consistency between the spectroscopic and photometric analysis. 

By integrating the observed SBP ($\mathrm{L}_{\rm T}$) and the three disk luminosity distributions ($\mathrm{L}_{\rm D}$) from the galactic center until \rcons, we computed \fd, the light fraction within the bulge residing in the disk under each of the assumptions:

\begin{equation}
\renewcommand*{\arraystretch}{1.5}
\begin{array}{l}

\mathrm{L}_{\rm D} = 2 \pi \int_{0}^{\mathrm{R}_{\rm C}} \mathrm{R}^{\star} \cdot 10^{(\mu_{D}(\mathrm{R}^{\star})-C)/-2.5} \cdot d \mathrm{R}^{\star}\\

\mathrm{L}_{\rm T} = 2 \pi \int_{0}^{\mathrm{R}_{\rm C}} \mathrm{R}^{\star} \cdot 10^{(\mu(\mathrm{R}^{\star})-C)/-2.5} \cdot d \mathrm{R}^{\star}\\
  
f_{D} =  \mathrm{L}_{\rm D} /  \mathrm{L}_{\rm T} \\
\end{array}
\renewcommand*{\arraystretch}{1}
\end{equation}

where $\mu(\rm R^{\star})$ is the surface brightness distribution of the observed SDSS $r$ or $g$-band SBP, $\mu_{\rm D}(\rm R^{\star})$ is the surface brightness distribution of the assumed disk component and $C$ the calibration constant.
\\

Bearing in mind the differences between the two data-sets (photometric and spectroscopic data), a mandatory step to achieve consistency when combining both techniques is to to scale the spectra based on the photometrically predicted luminosity fraction of the disk inside the bulge both in the SDSS $g$ and $r$ band. This is attained by convolving the bulge/disk spectra by the filter transmission curve \sdssT. Subsequent integration in the considered $\lambda$ range provides the corrected fluxes for the bulge $\mathrm{S}_{\rm C}$, and for each of the assumed disk luminosity distributions $\mathrm{S}_{\rm D}$:

\begin{equation}
\renewcommand*{\arraystretch}{1.5}
\begin{array}{l}

\mathrm{S}_{\rm D} = \int_{\lambda_{\rm min}}^{\lambda_{\rm max}}  \mathrm{\hat{F}}_{\rm D}^{\star}(\lambda) \cdot \mathrm{T}_{\rm SDSSg}(\lambda) \cdot d\lambda\\

\mathrm{S}_{\rm C} = \int_{\lambda_{\rm min}}^{\lambda_{\rm max}} \mathrm{F}_{\rm C}(\lambda) \cdot \mathrm{T}_{\rm SDSSg}(\lambda) \cdot d\lambda\\

\end{array}
\renewcommand*{\arraystretch}{1}
\end{equation}

Division of $\mathrm{S}_{\rm D}$ by $\mathrm{S}_{\rm C}$, after multiplication by the number of spaxels contained within the zones inside \rcons\	($n_{\rm pC}$) (\diskspec\ is normalized, i.e., it corresponds to a single spaxel) will result in the corrective factor, \fc:

\begin{equation}
f_{\rm C} =  n_{\rm pC} \cdot \mathrm{S}_{\rm D} / \mathrm{S}_{\rm C}
\end{equation}
 
The final scaling factor \fs, is simply the division between \fd, i.e., the light fraction within the bulge residing in the disk according to the surface photometry, and the corrective factor \fc, i.e., the same but according to the spectroscopic analysis, after correcting from the filter transmission curve:
 
\begin{equation}
f_{\rm S} = f_{\rm D} / f_{\rm C}
\end{equation}

The individual bulge net spectra for each of the three different disk configurations were computed by subtracting the scaled disk spectra from \bulgespec\ as:

\begin{equation}
    \mathrm{F}_{\rm B}(\lambda) = \mathrm{F}_{\rm C}(\lambda) - n_{\rm pC} \cdot f_{\rm S} \cdot \mathrm{\hat{F}}_{\rm D}^{\star}(\lambda)
\end{equation}
\\

Finally, the three $\mathrm{F}_{\rm B}(\lambda)$ (one for each disk configurations) for each galaxy were re-fitted by means of \starlight\ \& \fado\ with libraries Z4 \& Z5, this way completing the four spectral modeling runs.
\\

\subsection{Overview of the disk-subtraction tool:}
Summarizing, the developed suite of codes in Fortran, ESO-MIDAS and Python:

\begin{itemize}
  \item[i.] computes the integrated observed spectrum of the bulge -- OBS bulge, $\mathrm{F}_{\rm C}(\lambda)$, and the average spectrum of the disk -- OBS disk, \diskspecN;
  \item[ii.] by integrating the observed SBPs and those assumed for the disk, computes the scaling factors $f_{\rm S}$, under consideration of the SDSS $g$- or $r$-band filter transmission curves (for the $g$-band filter the average of the scaling factors of the sample are 2.37 for \brem{expD}, 2.09 for \brem{flatD} and 0.11 for \brem{decrD});
  \item[iii.] extracts and models with \starlight\ \& \fado\ the OBS bulge \& OBS disk, and determines the respective best-fitting stellar SEDs,  $\mathrm{F}_{\rm C}^{\star}(\lambda)$ \& $\mathrm{\hat{F}}_{\rm D}^{\star}(\lambda)$; 
  \item[iv.] scales and subtracts the latter from OBS bulge and re-fits (four times; i.e. with \starlight\ \& \fado\ for both the Z4 and Z5 SSP libraries) the net-bulge spectra $\mathrm{F}_{\rm B}(\lambda)$, obtained after subtraction of the three different functional forms of the disk within \rbulge, obtaining the best-fitting stellar SED for the net-bulge $\mathrm{F}_{\rm B}^{\star}(\lambda)$.
\end{itemize}

Figure~\ref{disk_sub_im} displays the results obtained for the LTG \object{IC 2604}. Extrapolation of a pure exponential or flat profile for the disk to the galactic center yields negative flux for $\lambda$ $\leq$ 4000~\AA, implying that the disk profile has to flatten or show a central depression within \rbulge. While assuming exponential and flat disk models, SLZ4's stellar age and metallicity estimates reached the maximum allowed value, which is per se an indication of an unphysical fit. Additionally, comparison of the resulting mass estimates for the bulge after subtraction of the exponential and flat models, $\rm \log({\cal M}_{\star, eD})$ and $\rm \log({\cal M}_{\star, fD})$, respectively, with the same prior of subtraction $\rm \log({\cal M}_{\star, B})$, indicates an increase of the stellar mass after disk removal, once more pointing to the invalidity of the assumed profiles for the disk (see next Sect.). For this galaxy, most of these criteria (used to classify the resulting net-bulge SED as unphysical) were met for all the four spectral modeling runs.

\section{Spectroscopic subtraction of the disk and first insights on the invalidity of its exponential intensity profile inside \rbulge}\label{resDiskSub}

The tool developed for the spectroscopic subtraction of the disk SED within \rbulge\ was applied to the entire sample using the previously estimated photometric constraints in both $r$- and $g$-band to scale the normalized spectrum of the disk. Considering that the results for both passbands are in agreement (within $\lesssim$ 15\%) and that the $g$-band transmission curve covers a significant part of the blue spectral range ($\sim$ 3830 - 5480 \AA), therefore better tracing the luminosity distribution of a SF disk, it was decided to present here only the results obtained from photometric decomposition $g$-band SBPs.


The first and possibly most decisive test of the validity of the  
exponential fitting function for the disk within \rbulge\ is to examine the properties of the residual net spectrum of the bulge after subtraction of the estimated contribution from the underlying disk, i.e.,  whether $\mathrm{F}_{\rm B}(\lambda)$ is positive throughout the considered spectral range. Since $\mathrm{F}_{\rm B}(\lambda)$ is obtained on the standard assumption that the disk exponential slope is preserved all the way to the center, a possibly negative flux of the net spectrum of the bulge over a significant spectral interval yields a strong indication against the validity of the background assumption of the disk exponentiality. Whereas this \emph{reductio ad absurdum} approach does not permit to directly constrain the intensity of the disk beneath the bulge, it allows to confirm or exclude a range of functional forms for the disk profile. Finally, we re-fitted the three $\mathrm{F}_{\rm B}(\lambda)$ with the purpose of evaluating the plausibility of the net-bulge SEDs, this way indirectly assessing the validity of the enclosed assumptions for the disk luminosity profile.

\begin{figure}[p]
\centering
\includegraphics[width=1.0\linewidth]{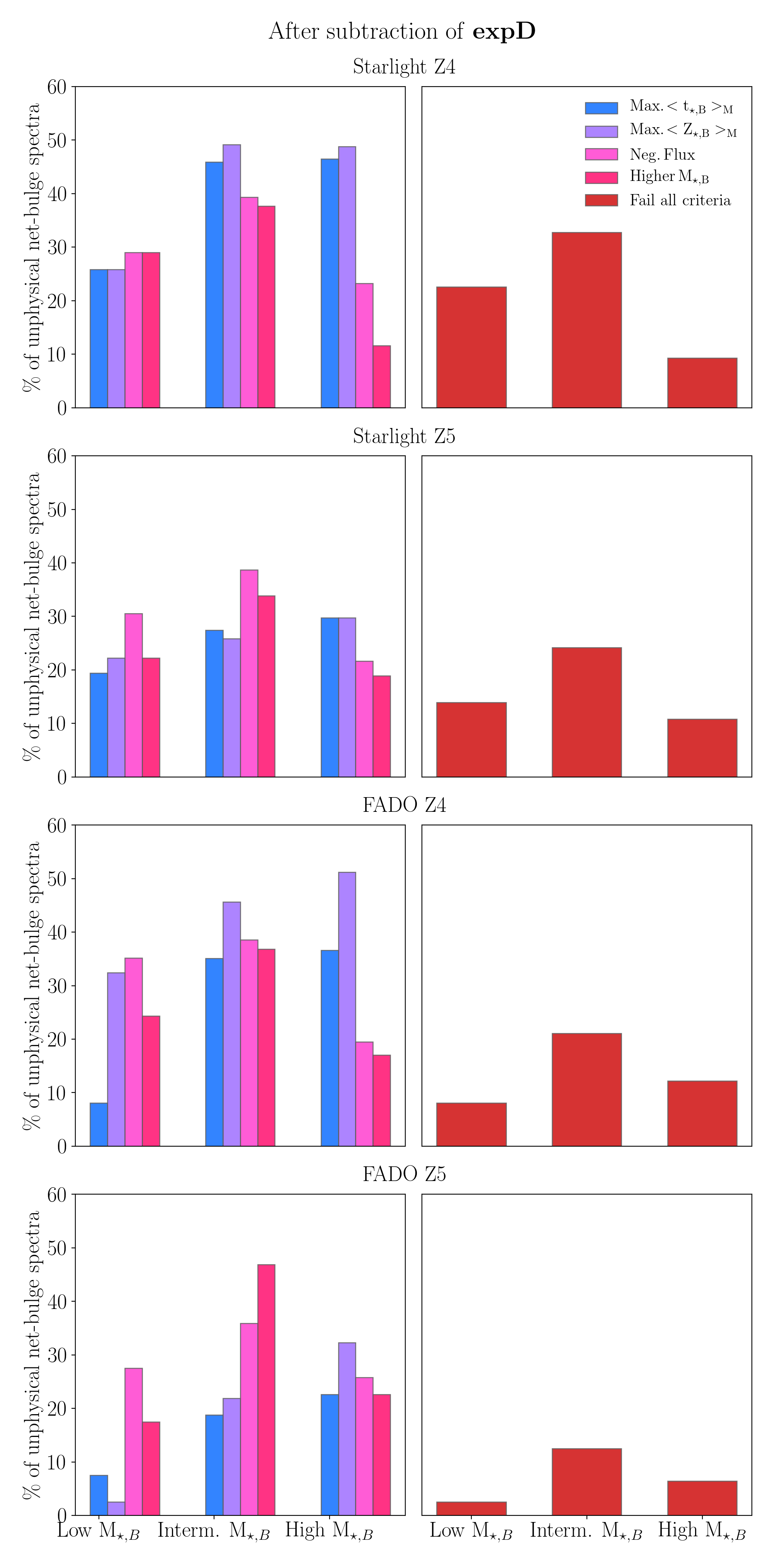}
\caption{Panels in the l.h.s. display the histograms showing the fractions of unphysical net-bulge spectra after subtraction of \brem{expD}, subdivided in three bulge mass bins (low $\rm M_{\star, B}$ for bulges with log of stellar mass lower or equal than 9.5 \msolar, intermediate $\rm M_{\star, B}$ for bulges with log of stellar mass between 9.5 and 10.5 and high $\rm M_{\star, B}$ for bulges with log of stellar mass higher or equal than 10.5). From left to right, the bars represent the fraction of net-bulge spectra that reached the maximum mass-weighted stellar age (blue) and metallicity (purple), display negative flux (light pink) and which estimated stellar mass is higher after disk subtraction (dark pink). Panels in the r.h.s contain the histograms showing the fractions of unphysical net-bulge spectra that meet all the aforementioned criteria. From top to bottom the rows refer to the spectral modeling run SLZ4, SLZ5, FDZ4 and FDZ5.}
\label{expD_prob}
\end{figure}

\begin{figure}[p]
\centering
\includegraphics[width=1.0\linewidth]{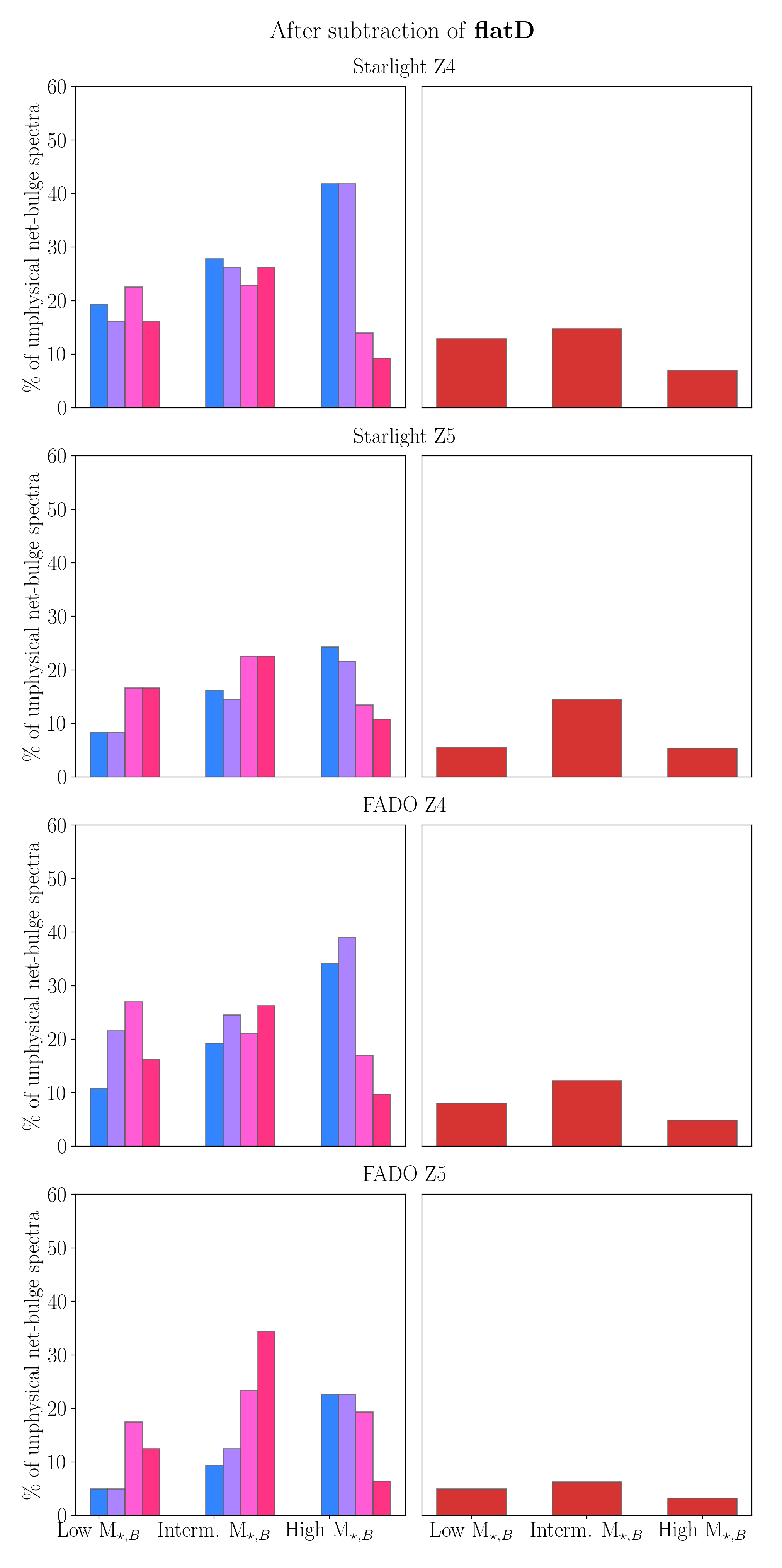}
\caption{Same layout as in Fig.~\ref{expD_prob}, displaying the results obtained after subtraction of \brem{flatD}.}
\label{flatD_prob}
\end{figure}

\begin{figure}[p]
\centering
\includegraphics[width=1.0\linewidth]{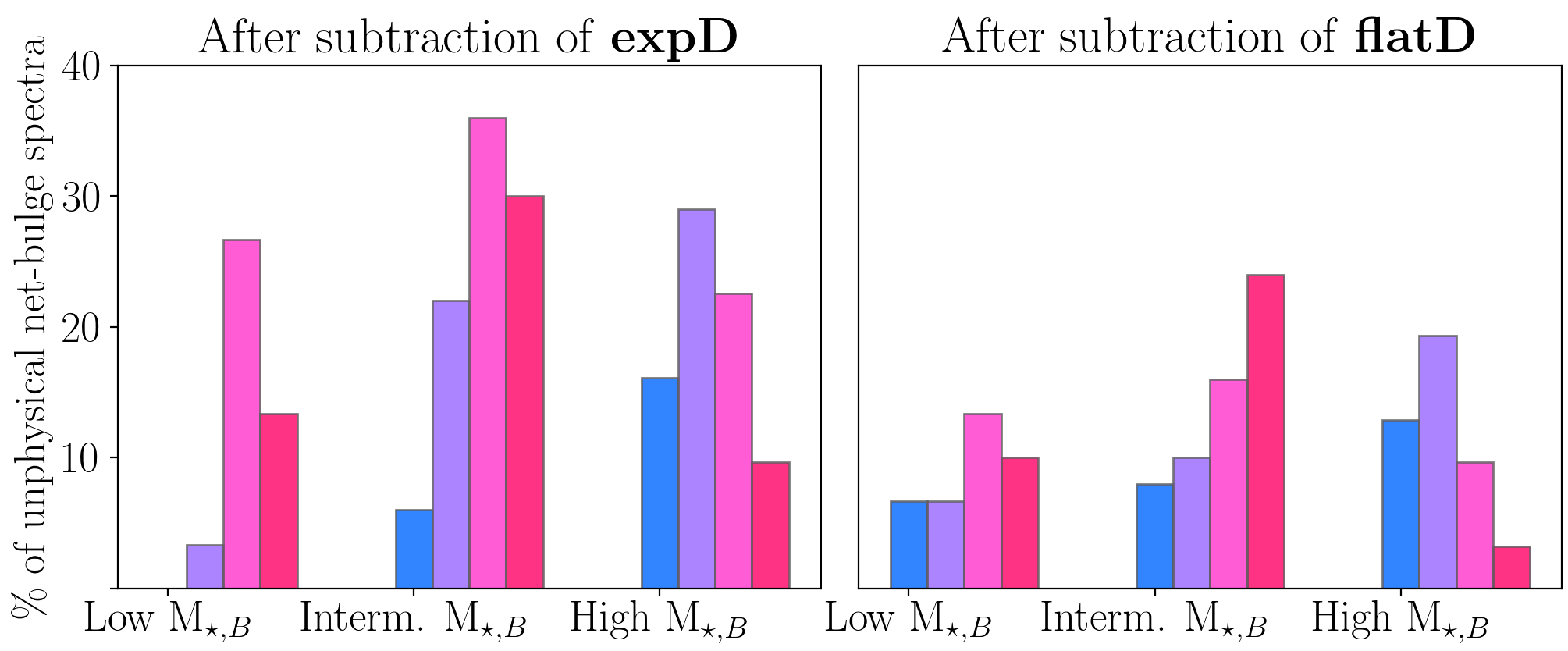}
\caption{Same color coding as in Figs.~\ref{expD_prob} \& \ref{flatD_prob}, showing the galaxy fractions failing the criteria simultaneously for all runs.}
\label{fail_all}
\end{figure}

\begin{figure}[p]
\centering
\includegraphics[width=1.0\linewidth]{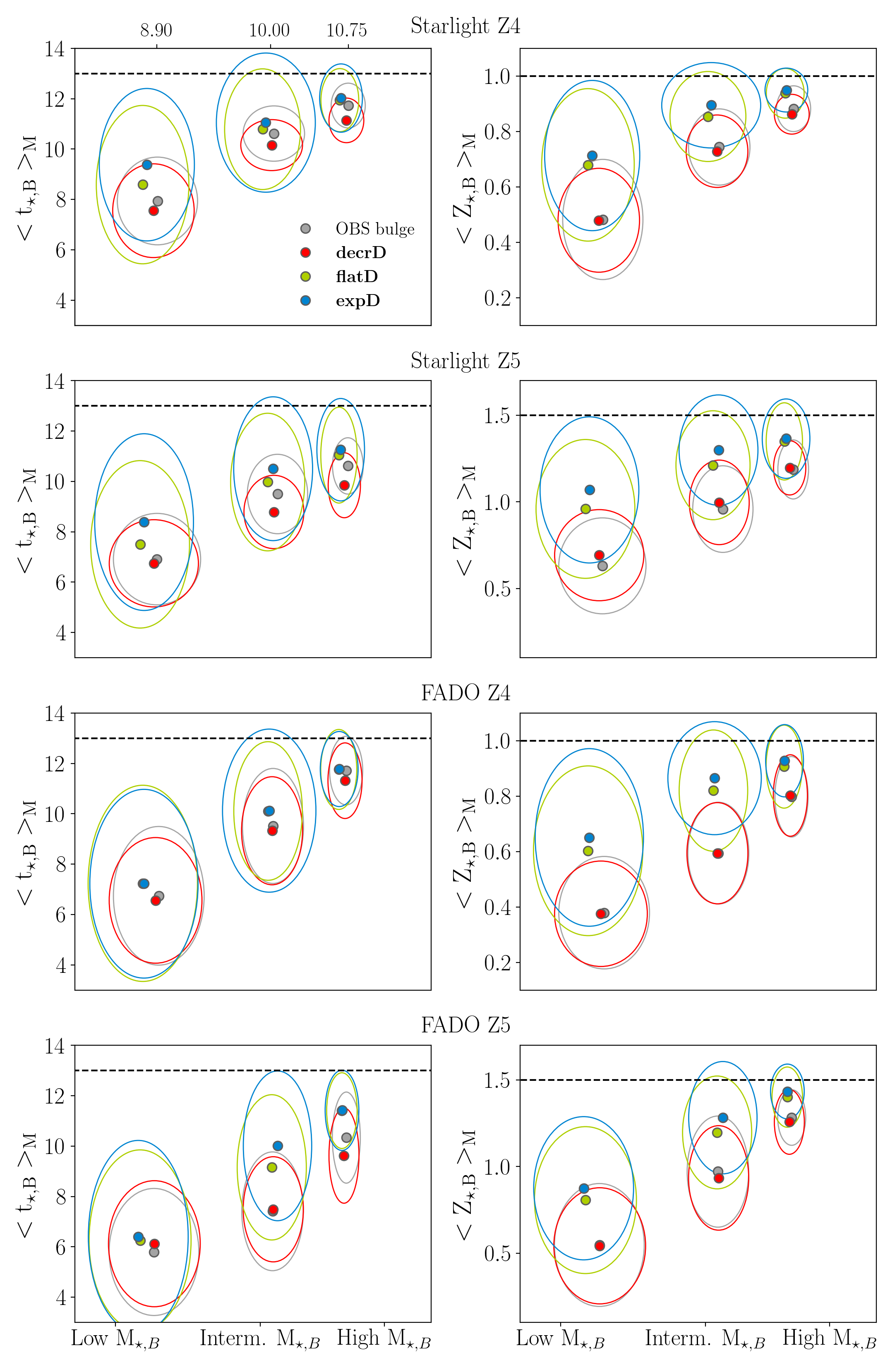}
\caption{
Panels in the l.h.s. display the mean values for the bulge's mass-weighted stellar age \mtmass\ in Gyr, in each mass bin, in the case of no subtraction (grey), subtraction of a centrally depressed disk (red), subtraction of centrally flattened disk (green) and subtraction of exponential disk (blue). Panels at the r.h.s. display the mean estimated values for the bulge's mass-weighted stellar metallicity \mZmass\ in \zsolar. The ellipses represent the standard deviation $\sigma$ of the mean, providing an estimate on how spread are these estimates within each bin mass. Their semi-major/minor axes display the $\sigma$ in $\rm M_{\star, B}$ (in the x-axis) and the $\sigma$ in \mtmass\ or \mZmass\ (in the y-axis). The 1$\rm ^{st}$ panel contains an additional x-axis providing information on the average stellar mass for each bulge bin mass. From top to bottom the rows refer to the four spectral modeling runs. The horizontal dashed line indicates the maximum allowed value for the respective quantity and SSP library.
}
\label{mean_values}
\end{figure}

Histograms displayed in Figs.~\ref{expD_prob} \&~\ref{flatD_prob} illustrate the percentage of unphysical net-bulge SED after subtraction of \brem{expD} and \brem{flatD}, respectively, in each of the three bulge stellar mass bins $\rm M_{\star, B}$ (the first bin encloses galaxies that host bulges with $\rm log(M_{\star, B}) \leq 9.5$ \msolar, being mainly composed by \brem{iA} galaxies, the second, bulges with $9.5 < \rm log(M_{\star, B}) < 10.5$ ($\sim$ \brem{iB}) and the third, bulges with $\rm log(M_{\star, B}) \geq 10.5$ ($\sim$ \brem{iC}) -- the average value for each mass bin is $10^{8.9}$ \msolar\ (31 galaxies), $10^{10}$ (61 galaxies) and $10^{10.75}$ (43 galaxies), respectively, as shown in the top-horizontal axis of the top-left panel of Fig.~\ref{mean_values}. 
In the l.h.s. are the panels showing the frequencies for each individual criteria (blue and purple bars express the percentage of galaxies that reached the maximum mass-weighted stellar age and metallicity allowed by the SSP library, respectively, light pink bars the fraction of partly negative net-bulge SEDs and dark pink bars the percentage of galaxies in each mass bin which stellar mass estimate is higher after disk subtraction) whereas the r.h.s. displays the fraction of galaxies that fail all criteria simultaneously.
In addition, Fig.~\ref{fail_all} shows the fraction of unphysical net-bulge SED according to SLZ4, SLZ5, FDZ4 \& FDZ5, i.e., the fraction of net-bulge spectra which fail the aforementioned criteria for all spectral modeling runs, simultaneously. As a complement, Table~\ref{table_sum} summarizes the fractions of unphysical net-bulge spectra obtained after subtraction of \brem{expD} \& \brem{flatD} relatively to the total galaxy sample.

Visual inspection of the light pink bars in the histograms show that there is no significant preference for any bulge-mass interval and that a significant fraction of bulges do not contain enough flux in their blue spectral range to accommodate the inwardly extrapolated disk profile. As a final thought on our barred galaxies, inspection of the true color images and $g$-$i$ color maps of our sample reveals that the frequency of bars is higher for massive galaxies and that generally bar colors are similar to those of the bulge, i.e., redder than those of the disk. Consequently, the possible inclusion of the bar in the normalized spectrum of the disk $\mathrm{F}_{\rm D}(\lambda)$ would lead to a slightly redder SED. Therefore, if contamination of $\mathrm{F}_{\rm D}(\lambda)$ by the bar was significant, one would expect negative (or very low) values also in the red spectral regime for some of the barred, high mass galaxies (where the bar contribution is significantly larger as compared to other mass bins). Such effect is not observed which lead us to conclude that, by adopting this methodology, bar contamination does not produce a significant effect.

Examination of the bars colored blue and purple show, for each bulge mass interval, the fraction of net-bulge spectra whose determined mass-weighted stellar age and metallicity has reached the maximum allowed value by the SSP library, respectively. Such \emph{failed} fits yield indirect constraints on the validity of the assumed model for the intensity profile of the disk within the bulge radius.
Inspection of these results demonstrates that, independently of the  
used PSS code or SSP library, a higher fraction of bulges in  
the higher mass bin tends to reach maximum values of age and  
metallicity, followed by the intermediate mass bulges and finally by low mass bulges, which display the lowest percentage of \emph{failed} fits, according to these criteria. Indeed, considering that the two higher mass bins enclose the intrinsically oldest and most metal-rich bulges in the sample, it is to be expected that mainly for these galaxies, subtraction of the rather blue spectrum of the disk will lead to a strong deficit of flux in the blue spectral range, thereby forcing PSS codes to reach maximum age\footnote{In addition to the typical uncertainties expected from spectral synthesis \citep[0.2-0.3 dex][]{Cid05,Cid14} resolving stellar populations becomes increasingly challenging with increasing stellar age. Based on \cite{Car19} who explore how \starlight\ and \fado\ recover the mass-weighted mean stellar age and metallicity and a set of additional tests performed adopting a similar experimental setup, we estimate the effective time resolution in age determinations of old stellar populations (>9 Gyr) to $\sim$1 Gyr.} and metallicity determinations.

This same effect is equally responsible for the increase in stellar mass after disk subtraction (dark pink bars), observed for a significant part of the galaxies of the sample: when fitting a spectrum using PSS codes, after obtaining the population vector PV (i.e, the fractional contributions of the individual SSPs), the stellar mass is derived by converting light to mass using the individual SSP's mass-to-light (M/L) ratios. Young stellar populations, which are typically seen in SF disks of LTGs, are characterized by low M/L ratios -- in spite of being significantly bright they contain a low percentage of stellar mass. On the other hand, older stellar populations, which typically populate the bulges of the most massive LTGs, have higher M/L ratios which imply that such stellar populations, although faint, constitute the bulk of the total stellar mass of the galaxy. Whereas it is common to observe an increase in flux, from the red to the blue spectral range, in the continuum of the spectra of SF disks, in the spectra of bulges is frequently observed a shortage of blue flux (i.e., a decrease in the continuum within the blue spectral range).
Depending on the steepness of the blue slope of the continuum of $\mathrm{F}_{\rm D}(\lambda)$ and on the deficiency of the flux of $\mathrm{F}_{\rm C}(\lambda)$ in this same spectral range, by removing the light contribution of a SF disk from the integrated central spectra, one might significantly reduce the blue flux of the residual SED. Seemingly, this approach is producing unreasonably red spectra (i.e., a severe lack of flux in the blue spectral range) for a significant fraction of the sample galaxies. In such cases, the PSS codes have no other alternative than to compensate this effect by introducing high fractions of high M/L SSPs (i.e., old stellar populations), which will artificially elevate the total stellar mass often to values which are higher than the ones derived from the observed integrated central spectrum. Logically, this result also implies that the assumptions retained in the adopted methodology are unreasonable, strengthening the conclusion that, for a significant part of the galaxies of the sample, the disk must diverge from exponentiality within \rbulge.

Figure~\ref{fail_all} displays the fractions of unphysical net-bulge SED within each bin mass for all spectral modeling run simultaneously. Its inspection and comparison with Figs.~\ref{expD_prob} \& ~\ref{flatD_prob} (see also Tab.~\ref{table_sum}) provides an idea of what are the criteria that are less impacted by the choice of PSS code or stellar library. Clearly and as expected, the most independent criteria are the frequency of net-bulge spectra with partially negative flux, which should be considered the most reliable test to define whether or not the enclosed assumptions are valid. Notwithstanding, even though the remaining criteria are more affected by the non-negligible uncertainties inherent to spectral synthesis (namely, for instance, age-metallicity degeneracy, different recipes for the convergence in distinct PSS codes, and increased difficulty in resolving old stellar populations) together they provide additional clues on the plausibility of the obtained net-bulge SEDs.

Considering that gradients of the stellar populations within the individual stellar components might be an important factor determining the unplausibility of the obtained net-bulge SEDs, we explored $g$-$r$ color and stellar age gradients within bulge and disk.
Although color and age gradients within \rbulge\ might be quite considerable \citep[see][]{Bre20a}, the same are generally negligible for the disk component. Figure 2 of \cite{Bre20a} shows that intermediate stellar mass galaxies, which host intermediate mass bulges, generally do not display  significant age gradients. Bearing in mind that age gradients within the disk region are often insignificant and that the fractions of the net-bulge SEDs which are partially negative are higher for this particular mass bin, age/color gradients cannot be the main reason for the high number of partially negative net-bulge spectra.

\begin{table*}
\centering
\begin{tabular}{ c c c c c c }
 Run & Max. \mtmass\ & Max. \mZmass\ & Neg. Flux & > $\rm M_{\star, B}$ & All criteria \\ 
\hline  
\multicolumn{6}{c}{After subtraction of \brem{expD}} \\
\hline  
SLZ4 & 41.5 & 43.7 & 31.9 & 27.4 & 23.0 \\
SLZ5 & 25.9 & 25.9 & 31.9 & 26.7 & 17.8 \\
FDZ4 & 28.1 & 43.7 & 31.9 & 27.4 & 14.8 \\
FDZ5 & 16.3 & 18.5 & 31.1 & 32.6 & 8.1 \\
\hline  
Average & 28.0 & 33.0 & 31.7 & 28.5 &  15.9 \\
\hline  
\multicolumn{6}{c}{After subtraction of \brem{flatD}} \\
\hline  
SLZ4 & 30.4 & 28.9 & 20.0 & 18.5 & 11.9 \\
SLZ5 & 16.3 & 14.8 & 18.5 & 17.8 & 9.6 \\
FDZ4 & 21.5 & 28.1 & 21.5 & 18.5 & 8.9 \\
FDZ5 & 11.1 & 12.6 & 20.7 & 21.5 & 5.2 \\
\hline  
Average & 19.8 & 21.1 & 20.2 & 19.1 & 8.9 \\
\end{tabular}
\caption{Percentage of unphysical net-bulge SEDs after subtraction of disk models \brem{expD} and \brem{flatD} relatively to the total number of galaxies in the sample.}\label{table_sum}
\end{table*}

In addition, Fig.~\ref{mean_values} displays the average values for the bulge mass-weighted mean stellar age \mtmass\ (l.h.s.), and metallicity \mZmass\ (r.h.s.), within each mass bin, as obtained for the four spectral modeling runs. Major/minor axes of the ellipses depict the error bars ($\sigma$ of the mean) for the estimated stellar mass (horizontal axes) and for \mtmass\ or \mZmass\ (vertical axes). The different colors correspond to the various assumptions that were tested  -- grey dots depict the mean values for OBS bulge $\mathrm{F}_{\rm C}(\lambda)$, red dots the values obtained after subtraction of the centrally depressed disk \brem{decrD}, green dots the values obtained after subtraction of the flattened disk \brem{flatD} and blue dots the values obtained after subtraction of the standard exponential disk \brem{expD}. Inspection of this Fig. and Tab.~\ref{table_sum} reveals that:

\begin{itemize}
   \item[i.] mass-weighted age and metallicity determinations are not  
significantly affected after subtraction of a centrally depressed disk (variations in mass, age or metallicity are within the expected error associated with spectral synthesis). Such a result was expected, considering that by assuming a disk shape such as \brem{decrD} the scaling factor \fs, i.e., the light fraction of the disk within \rcons, is in all cases low, with an average value of 11$\%$ for the whole sample.
   \item[ii.] generally, ages and metallicities obtained from fitting $\mathrm{F}_{\rm B}(\lambda)$ increase as compared to those prior to disk subtraction;
   \item[iii.] higher mass galaxies (which host higher mass bulges) have an increased tendency to reach the maximum value allowed by the SSP library; 
   \item[iv.] the higher the assumed disk contribution inside  
\rbulge\ (from \brem{decrD} to \brem{flatD} to \brem{expD}), the  
higher is the fraction of LTGs for which the determined age and/or  
metallicity converges to the maximum allowed value by the adopted SSP  
library (0$\%$ of the galaxies of the sample for \brem{decrD}, an average of $\sim$20 (21)$\%$ for \mtmass\ (\mZmass) after \brem{flatD} and an average of $\sim$28 (33)$\%$ after \brem{expD}). The same behavior is observed for the increase in stellar mass after disk subtraction (0$\%$ after \brem{decrD}, an average of $\sim$20$\%$ after \brem{flatD} and $\sim$28$\%$ after \brem{expD}) and fraction of partially negative SEDs (0$\%$ after \brem{decrD}, an average of $\sim$20$\%$ after \brem{flatD} and $\sim$32$\%$ after \brem{expD});
   \item[vi.] the values obtained with \fado\ are in every case more disperse as compared to the same obtained with \starlight\ (even for OBS bulge and \brem{decrD}) evidencing the non-negligible discrepancies in quantities obtained by different spectral synthesis codes.
\end{itemize}

%





Although instructive, this exercise alone is not sufficient to definitively answer the question of whether the disk preserves its intensity slope, or even exists, inside the bulge radius. 
Nevertheless, this investigation has placed important constraints on the possible light distribution by the disk within \rbulge: for a substantial part of the sampled LTGs, independently of their stellar mass,
the assumption of inward extrapolation of the exponential intensity  
profile of the disk yields dubious or unphysical results. This suggests that the  
true intensity profile of the disk inside \rbulge\ shows a flattening  
or central depression, as proposed from theoretical studies. If, on the other hand, the disk light distribution within \rbulge\ conserves its exponential nature, these results indicate that, for a significant fraction of the analyzed galaxies, the disk contribution within \rr\ $<$ \rbulge\ should be much redder than the host disk, displaying a spectroscopic profile more similar to the bulge.

\section{Summary and Conclusions}\label{conc}

With the goal of placing constraints on the radial intensity profile of the disk in late-type galaxies within their bulge radius \rbulge, we developed a tool that allows us to determine the net SED of the bulge 
after spectroscopic subtraction of the photometrically inferred contribution from the underlying disk. Although quite rudimentary at this stage, this technique allowed us to gain first insights into the validity of the standard assumption that the disk preserves its exponential slope all the way to the LTG center:
\begin{itemize}
   \item[i.] The analysis presented here indicates that, independently of the bulge's stellar mass (which is tightly correlated with the total LTG stellar mass and its bulge's mean stellar age), up to 32\% (20\%) of the SEDs obtained for the bulge after subtraction of an exponential (or inwardly flattening) model for the disk
   yield negative flux in the blue spectral range. This implies that  
in a significant fraction of LTGs the disk component must show a  
central flat core or intensity depression inside \rbulge. 
   \item[ii.] Further support against the standard assumption of the  
exponentiality of the disk within \rbulge\ comes from the fact that
   spectral modeling obtained in that case for the net SED of the  
bulge leads to dubious results. Specifically we find that when a purely exponential disk profile is assumed, $\sim$28\% ($\sim$33\%) of the disk-subtracted bulges reach the maximum allowed age (metallicity), for $\sim$28\% the stellar mass is higher than that estimated prior to disk subtraction and for $\sim$32\% the remaining net-bulge SED were partially negative.
     \item[iii.] By assuming a flat intensity profile for the disk  
within \rbulge, spectral modeling of $\sim$20\% ($\sim$21\%) of the net SED of bulges in our sample yields the maximum allowed age (metallicity), for $\sim$ 20\% of all cases the stellar mass exceeds that estimated  
within \rbulge\ prior to disk subtraction and for $\sim$20\% the remaining net-bulge SED were partially negative.
\end{itemize}

The present investigation suggests that, in a significant fraction of LTGs, the disk component must show a central intensity depression inside \rbulge. If proven to be true, the soundness of the outcome resulting from a substantial fraction of past studies would be compromised, namely the ones based on bulge/disk decomposition, considering that this issue would propagate to many local and moderate to high-$z$ studies, impacting, for instance findings concerning growth and evolution of bulges and disks of LTGs (e.g., the evolution in $z$ of the B/T ratio).  
Facing these results, one can even speculate that some of the reported findings regarding the formation and evolution of LTGs (see Sect.\ref{intro}) might be artificially driven by the wrong assumption of the conservation of the exponential disk within the bulge while performing structural decomposition.
Furthermore, it is worth bearing in mind that, in disk-dominated LTGs, the assumed (inwardly extrapolated) disk flux can provide up to $\sim$ 80\% of the light inside the bulge radius. Therefore, overestimating the true luminosity fraction by the disk could lead to the erroneous classification of a high-luminosity, high-$\eta$ CB as a PB. Moreover, if this bulge hosts an active galactic nucleus (AGN) one will conclude that some PBs display AGN activity in their cores whereas in fact this bulge is not a PB but a CB. This hypothetic scenario serves only as an example of how the adopted assumptions and methods might dictate the obtained results.

To complement this analysis and further explore the issue one can take advantage of the excellent-quality IFS data captured by 10m-class telescopes (e.g., with the MUSE@VLT spectrograph) being now at hand. By performing spatially resolved spectral synthesis to a number of local, moderately inclined LTGs one could explore more deeply the radial mass and stellar surface density profiles of galactic disks, based on this spectrophotometric decomposition technique. Moreover, the high spectral resolution and S/N permitted by MUSE data permits to resolve older stellar populations significantly more accurately as compared to IFS data restricted to the blue spectral range, especially when using a self-consistent spectral modeling tool, such as \fado.
Finally, via kinematical decomposition, one could investigate $\rm V_{rot} /	\sigma_{\star}$ radial profiles, which might too give further insights on the validity of the inward exponentiallity of galactic disks.

\begin{acknowledgements}
We thank the anonymous referee for valuable comments and suggestions.  This work was supported by Fundação para a Ciência e a Tecnologia (FCT)
through the research grants [UID/FIS/04434/2019] UIDB/04434/2020 and
UIDP/04434/2020.
I.B. was supported by Instituto de Astrof\'isica e Ci\^encias do Espa\c{c}o through the research grant CIAAUP-30/2019-BID and by the FCT PhD::SPACE Doctoral Network (PD/00040/2012) through the fellowship PD/BD/52707/2014 
funded by FCT (Portugal).
P.P. was supported through Investigador FCT contract IF/01220/2013/CP1191/CT0002 and by a contract that is supported by FCT/MCTES through national funds (PIDDAC) and by grant PTDC/FIS-AST/29245/2017.
J.M.G. is supported by the DL 57/2016/CP1364/CT0003 contract and
acknowledges the previous support by the fellowships CIAAUP-04/2016-BPD in
the context of the FCT project UID/-
FIS/04434/2013 \& POCI-01-0145-FEDER-007672, and SFRH/BPD/66958/2009 funded by FCT and POPH/FSE (EC).
This study uses data provided by the Calar Alto Legacy Integral Field Area (CALIFA) survey (califa.caha.es), 
funded by the Spanish Ministry of Science under grant ICTS-2009-10, and the Centro Astron\'omico Hispano-Alem\'an.
It is based on observations collected at the Centro Astron\'omico Hispano Alem\'an (CAHA) at Calar Alto, operated jointly 
by the Max-Planck-Institut f\"ur Astronomie and the Instituto de Astrof\'isica de Andaluc\'ia (CSIC).
This research has made use of the NASA/IPAC Extragalactic Database (NED) which is operated by the Jet Propulsion Laboratory, 
California Institute of Technology, under contract with the National Aeronautics and Space Administration.
\end{acknowledgements}




\end{document}